\documentclass[a4paper,pdftex,rmp,twocolumn,amsmath,amssymb,groupedaddress,floatfix]{revtex4}

\usepackage{float}
\usepackage{graphicx}
\usepackage[dvipsnames]{xcolor}
\usepackage{accents}
\usepackage{hyperref}
\usepackage[normalem]{ulem}
\usepackage{enumitem}
\usepackage{orcidlink}

\usepackage{amssymb}
\usepackage{amsmath}
\usepackage{accents}
\usepackage{mhchem}
\setcitestyle{super}

\newcommand{\be}{\begin{equation}}
\newcommand{\ee}{\end{equation}}
\newcommand{\bea}{\begin{eqnarray}}
\newcommand{\eea}{\end{eqnarray}}

\begin{document}
\title{Exploring 3D community inconsistency in human chromosome contact networks}

\date{\today}

\author{Dolores Bernenko\orcidlink{0000-0002-6618-8232}}
\email{dolores.bernenko@umu.se}
\affiliation{Department of Physics, Integrated Science Lab, Ume\r{a} University, SE-901 87 Ume\r{a}, Sweden}
\author{Sang Hoon Lee\orcidlink{0000-0003-3079-5679}}
\email{lshlj82@gnu.ac.kr}
\affiliation{Department of Physics and Research Institute of Natural Science, Gyeongsang National University, Jinju 52828, Korea}
\affiliation{Future Convergence Technology Research Institute, Gyeongsang National University, Jinju 52849, Korea}
\author{Ludvig Lizana\orcidlink{0000-0003-3174-8145}}
\email{ludvig.lizana@umu.se}
\affiliation{Integrated Science Lab, Department of Physics, Ume\r{a} University, SE-901 87 Ume\r{a}, Sweden}

\begin{abstract}

Researchers have developed chromosome capture methods such as Hi-C to better understand DNA's 3D folding in nuclei. The Hi-C method captures contact frequencies between DNA segment pairs across the genome. When analyzing Hi-C data sets, it is common to group these pairs using standard bioinformatics methods (e.g., PCA). Other approaches handle Hi-C data as weighted networks, where connected node pairs represent DNA segments in 3D proximity. In this representation, one can leverage community detection techniques developed in complex network theory to group nodes into mesoscale communities containing nodes with similar connection patterns. While there are several successful attempts to analyze Hi-C data in this way, it is common to report and study the most typical community structure. But in reality, there are often several valid candidates. Therefore, depending on algorithm design, different community detection methods focusing on slightly different connectivity features may have differing views on the ideal node groupings. In fact, even the same community detection method may yield different results if using a stochastic algorithm. This ambiguity is fundamental to community detection and shared by most complex networks whenever interactions span all scales in the network. This is known as community inconsistency. This paper explores this inconsistency of 3D communities in Hi-C data for all human chromosomes. We base our analysis on two inconsistency metrics, one local and one global, and quantify the network scales where the community separation is most variable. For example, we find that TADs are less reliable than A/B compartments and that nodes with highly variable node-community memberships are associated with open chromatin. Overall, our study provides a helpful framework for data-driven researchers and increases awareness of some inherent challenges when clustering Hi-C data into 3D communities.

\end{abstract}

\maketitle

\section{Introduction}

Chromosomes' three-dimensional (3D) folded structure is critical to understanding genetic processes and genome evolution. The discovery of these 3D structures relied on the analysis of pan-genome-wide chromosome capture data, represented by a pairwise interaction matrix called Hi-C map~\cite{lieberman2009comprehensive,dixon2012topological,rao20143d}. These maps reveal substructures of different scales, including the dichotomous division into A (active) versus B (inactive) compartments, determined from principal component analysis (PCA)\cite{lieberman2009comprehensive}, and smaller-scale topologically associated domains (TADs) identified using the Arrowhead algorithm\cite{rao20143d}. Several molecular biologists find these structures appealing because they define DNA regions with correlated gene expression and epigenetic modifications. In addition, their borders enrich binding sites for architectural proteins such as CTCF (11-zinc finger protein CCCTC-binding factor) in humans and CP190 in \emph{Drosophila melanogaster}\cite{magana2020chromatin}.

As researchers delved deeper into this crucial topic, they discovered that TADs act as shielded 3D domains with more internal than external contacts, similar to the definition of ``communities'' in network science~\cite{NewmanBook, Porter1, graph_detection}, and that some TADs are nested or partially overlapped. Also, studies of A/B compartments showed cross-scale organization as they split into six subclasses (A1, A2, B1, etc.)\cite{rao20143d, Sarnataro2017}. Furthermore, the authors of this paper developed a network-community-detection method that considers the average contact frequency between distant DNA segments based on their sequence distance. This method revealed a spectrum of mesoscale 3D communities in Hi-C data\cite{lee2019mapping, Bernenko2022}, ranging from A/B compartments to TADs. All these findings point to chromosomes as having a complex multi-scale structure.

Like Hi-C networks, most complex networks have a blend of overlapping communities at different scales, making community detection challenging. This complexity complicates finding statistically significant communities, as most community detection methods rely on the objective function called modularity~\cite{Porter1,graph_detection}. In this function, the scale is defined as the value of a ``resolution parameter''~\cite{resolution}, and most community detection algorithms select its value rather arbitrarily without a principled guideline (see Ref.~\cite{Newman2016} for the relationship between the resolution parameter and a parameter from another established community detection framework called the stochastic block model). Recently, some researchers (including one of the authors of this paper) have directly acknowledged this fact and tried extracting informative structural properties based on the ensemble of cross-scale ``inconsistently'' detected communities~\cite{kim2019relational,lee2021consistency,riolo2020consistency}.

This paper utilizes two representative inconsistency metrics~\cite{lee2021consistency}, one local and one global, to quantitatively assess the scales that provide the most reliable 3D communities in the Hi-C data. The reliability is based on the multi-scale ``landscape'' of Hi-C communities. In Sec.\ref{sec:methods}, we present the Hi-C data and the associated weighted networks. We also describe our inconsistency analysis framework and community detection method, and how we assign chromatin states to each Hi-C bin by calculating folds of enrichment ratios. In Sec.\ref{sec:results}, we present our findings and  conclude our paper in Sec.~\ref{sec:conclusions}.

\section{Methods}
\label{sec:methods}

\subsection{Transforming Hi-C data as weighted network}
We use the same Hi-C intra-chromosomal contact map as our previous series of studies~\cite{lee2019mapping,Bernenko2022} [human cell line GM12878 (B-lymphoblastoid)~\cite{rao20143d,edgar2002gene}]. Also, as before~\cite{lee2019mapping,Bernenko2022}, we use the MAPQG0 data set at the 100 kilobase-pair (kb) resolution and normalize the interaction map with the Knight-Ruiz (KR) matrix balancing~\cite{KR_norm}. As a result, we treat each 100 kb chromatin locus as the minimal unit, or ``node'', and the normalized interaction weights between nodes $i$ and $j$ as weighted edges, using network science terminology~\cite{NewmanBook}.

\subsection{Network community detection and inconsistency} 
One of the most popular ways to detect network communities~\cite{Porter1,graph_detection}---densely-connected substructures---is to maximize the objective function called modularity\footnote{We emphasize that we are using the terminology for weighted networks since we use the weighted version of the Hi-C interaction map. For binary networks, they are reduced to the conventional version: $A_{ij}$ is either $0$ or $1$ representing absence or presence of the edge, and $k_i$ is the number of neighbouring nodes to $i$ (the degree).}
\begin{equation}
    \mathcal{M} = \frac{1}{2m} \sum_{i \ne j} \left[ \left( A_{ij} - \gamma P_{ij} \right) \delta(g_i,g_j) \right ] \,.
\label{eq:modularity}
\end{equation}
Here, $A_{ij}$ denotes the adjacency matrix elements corresponding to the interaction weights between nodes $i$ and $j$ ($A_{ij}=0$ indicates no edge), and $P_{ij}$ is the expected edge weight based on \emph{a priori} information. The most popular choice is when considering only the overall tendency of node-node interaction $P_{ij} = k_i k_j / (2m)$, where $k_i$ represents node $i$'s strength (the sum of its weights) and $m$ is a normalization constant ensuring that $-1 \le \mathcal{M} \le 1$. Finally,  $g_i$ is the community index of node $i$ and $\delta$ is the Kronecker delta. A key parameter in our study is the resolution parameter $\gamma$, which controls the overall community scale~\cite{resolution}.

In principle, maximizing the modularity function with respect to all of the possible community divisions, encoded as $\{ g_i \}$ in Eq.~\eqref{eq:modularity}, is a mathematically well-defined deterministic concept. However, due to the computational limitation imposed by the problem, it is prohibitively difficult to find the exact solution, e.g., from the comprehensive enumeration of the network divisions. Therefore, most network community detection algorithms rely on various types of approximations or parameter restrictions. Many algorithms take a stochastic approach to sample the community partitions,  just as in standard Monte Carlo~\cite{MonteCarloBook}. One example is the Louvain-type algorithms~\cite{Louvain,GenLouvain} we use here (detailed in Sec.~\ref{sec:GenLouvain}). 
 
\begin{figure}
     \includegraphics[width = \columnwidth]{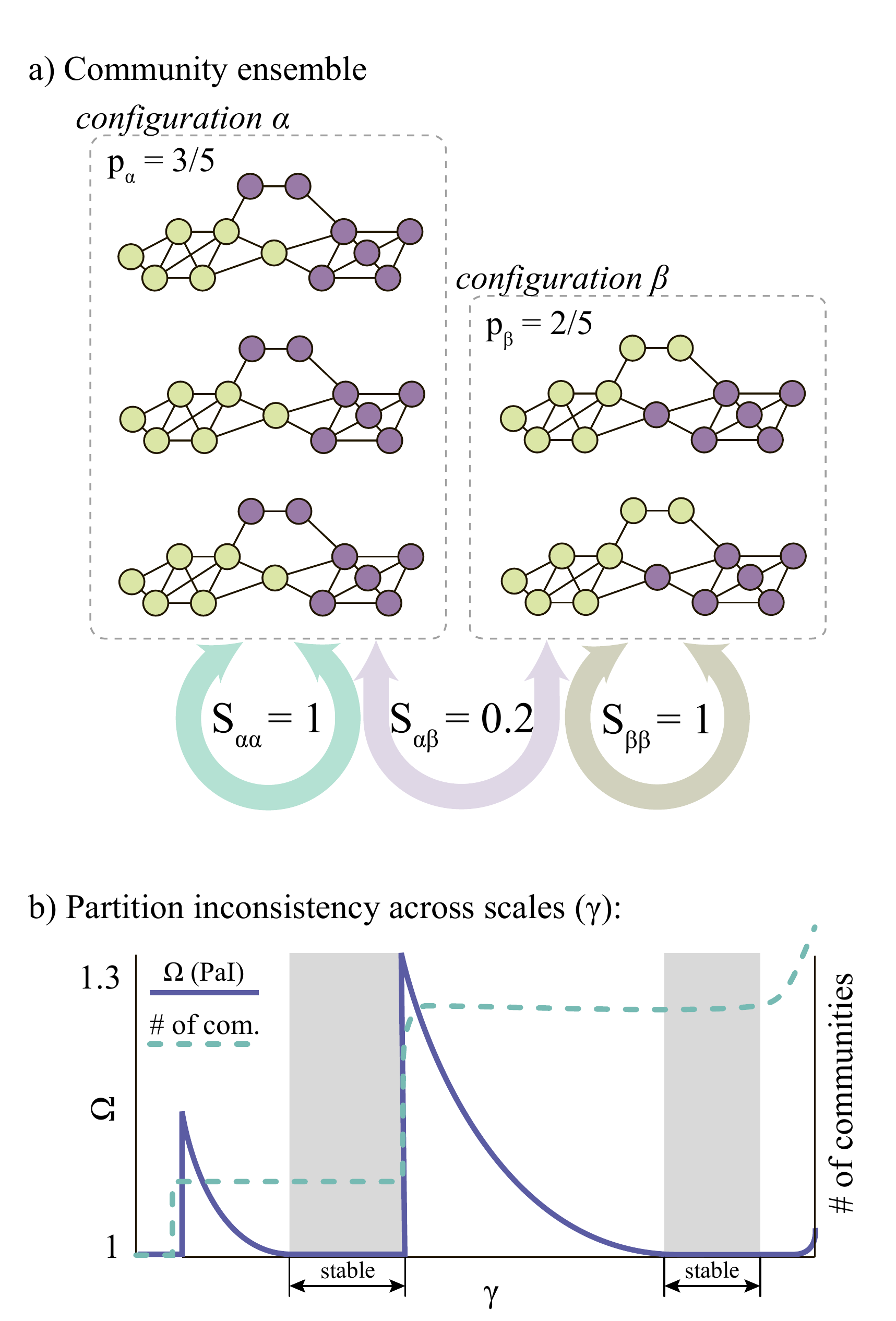}
     \caption{Illustration of partition inconsistency (PaI). (a) A community ensemble composed of two configurations $\alpha$ and $\beta$. The similarity measure~\cite{EC} quantifies the degree of similarity between the community configurations. Based on this measure, we calculate the PaI value $\Omega$ as in Ref.~\cite{lee2021consistency}. (b) Cross-scale PaI curve (varying resolution parameter $\gamma$) alongside the average number of communities. Shaded areas show ranges of statistically reliable community scales.}
     \label{fig:PaI}
 \end{figure}

\begin{figure}
     \includegraphics[width = \columnwidth]{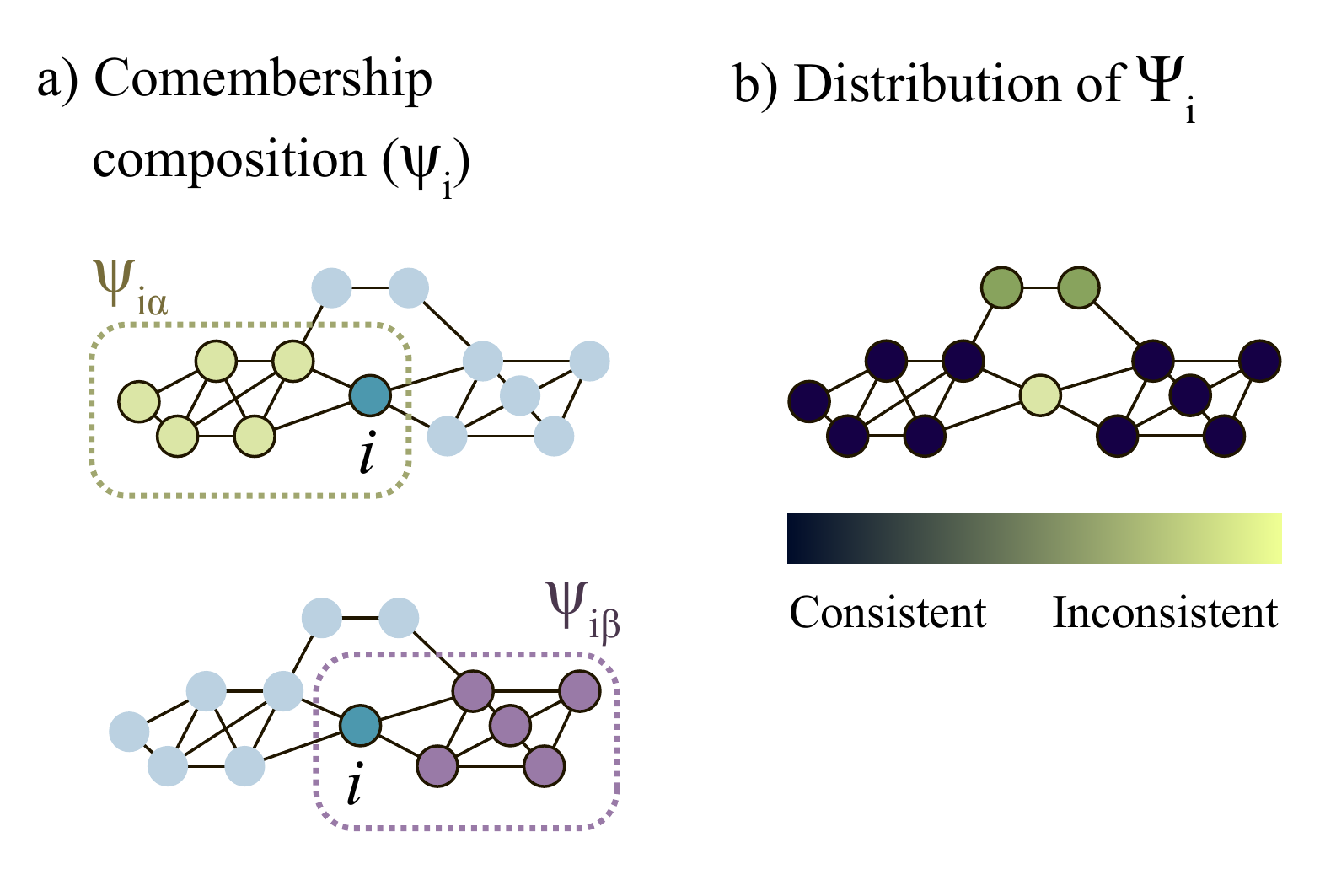}
     \caption{Illustration of the membership inconsistency (MeI). (a) The co-membership composition for node $i$ in configurations $\alpha$ and $\beta$. From this co-membership structure, we calculate the MeI value $\Psi_i$ as in Ref.~\cite{lee2021consistency}. (b) Example distribution of MeI for a small network.}
     \label{fig:MeI}
 \end{figure}

Although stochastic approaches like Louvain have been successful in terms of speed and accuracy in many  community detection applications, their stochastic nature may produce \emph{multiple} results that sometimes include \emph{inconsistent} elements. Researchers tend to work around this inconsistency~\cite{CC2,CC1} by choosing the most consistent, or reproducible, network partition. However, one of the authors of this paper has turned this inconsistency into an advantage, using it to probe network structural information~\cite{kim2019relational,lee2021consistency} at both global  and local levels (Figs.~\ref{fig:PaI} and \ref{fig:MeI}). In particular, by studying inconsistency measures one may pinpoint scale regimes or specific node collections that are the most statistically reliable (at the global level) or flexible (at the local level). For a detailed theoretical framework, we defer to Ref.~\cite{lee2021consistency}. But below, we remind the reader of the essential parts used in this analysis. 

One metric we study is partition inconsistency (PaI). It quantifies the global degree of inconsistency among community partitions in the entire network (Fig.~\ref{fig:PaI}). PaI is based on a recently developed similarity measure~\cite{EC} $S_{\alpha\beta}$ between community configurations $\alpha$ and $\beta$. The PaI value $\Omega (\ge 1)$ indicates the effective number of independent configurations. A small (large) $\Omega$ value represents more consistent (inconsistent) regimes, respectively. Using PaI, one extracts the most statistically reliable ranges of community scales by focusing on the (local) minima of $\Omega$, in particular, alongside another meaningful evidence of stable communities: the number of communities stays flat at a specific integer value [illustrated in Fig.~\ref{fig:PaI}b)]. 

While PaI describes the network's global inconsistency, we use another metric, membership inconsistency (MeI), to quantify local (individual-node) inconsistencies (Fig.~\ref{fig:MeI}). MeI represents the effective number of independent communities for a specific node across different community configurations. As shown in Fig.~\ref{fig:MeI}b), the MeI values properly detect the functionally flexible or ``bridge'' nodes participating in different modules\footnote{The MeI measure introduced in Ref.~\cite{lee2021consistency} is a more principled and improved measure than the original ``companionship inconsistency (CoI)'' measure first introduced in Ref.~\cite{kim2019relational}, by considering the possibility of more than two community memberships.}. In Sec.~\ref{sec:results}, we use PaI and MeI to study global and local community inconsistencies of Hi-C maps and relate to these metrics to other biological data.

\subsection{GenLouvain method}
\label{sec:GenLouvain}
The stochastic community detection method we utilize throughout our work is version 2.1 of GenLouvain~\cite{GenLouvain} (see \url{https://github.com/GenLouvain/GenLouvain} for the latest version). GenLouvain is a variant of the celebrated Louvain algorithm~\cite{Louvain}, which is one of the most widely used algorithms and is popular due to its speed and established packages in various programming languages. Starting from single-node communities, the algorithm accepts or rejects trial merging processes based on the modularity change in a greedy fashion.  To determine the community stability across network scales, we run GenLouvain several times, at least 100,  for each resolution parameter $\gamma$ and then calculate the global (PaI) and local (MeI) inconsistency metrics.

\subsection{Cross-scale node-membership correlations}
\label{sec:t_SNE}
We use GenLouvain to produce an ensemble of community partitions from Hi-C data, for fixed scale parameters $\gamma$. However, some of these partitions seems correlated. To better understand these correlations, we use a graphical embedding technique designed to illustrate high-dimensional data on a 2D plane. Specifically, we use t-SNE (t-distributed stochastic neighbor embedding)\cite{maaten_visualizing_2008}.

t-SNE is a general framework that aggregates data points based on some distance metric. While there are several choices, we use the so-called correlation distance $D$, which is common for random vectors and defined as 
\begin{equation}
     D = 1 - r(\mathbf u, \mathbf v),
    \label{eq:d_corr}   
\end{equation}
where $r(\mathbf u,\mathbf v)$ is the correlation between the vectors $\mathbf u$ and $\mathbf v$, conventionally defined as
\begin{equation}
     r(\mathbf u, \mathbf v) = \frac{(\mathbf u - \bar{\mathbf u}) \cdot (\mathbf v - \bar{\mathbf v})}  {{||(\mathbf u - \bar{\mathbf u})||}_2 {||(\mathbf v - \bar{\mathbf v})||}_2},  
\end{equation}
where $\bar{u}$ and $\bar{v}$ are  the mean of the elements (so that $\bar{\mathbf u} = \bar{u} \times [1,1,1,\ldots]$) and $|| \cdots ||_2$ is the Euclidean norm. According to Eq.~\eqref{eq:d_corr}, $D=0$ if they are perfectly correlated  ($r=1)$  and $D \approx 1$ if they are uncorrelated ($r \approx 0$).

In our analysis, we create these vectors from 100 GenLouvain runs. Each vector is a Boolean representation of the node-community membership for one node (each element is either $1$ or $0$) depending on whether the node belongs to a specific community at a particular GenLouvain iteration.
For our analysis, we used scikit-learn\cite{scikit-learn}. To reach the best visualization results, we tuned several parameters : perplexity: 20, early exaggeration: 8,initialization: random, and number of iterations: 1356.

\subsection{Chromatin states and enrichment}
\label{sec:chromatin_states}
In Results (Sec.~\ref{sec:results}), we analyze the inconsistency measures PaI and MeI in terms of chromatin states derived from an established chromatin division~\cite{hmm2011article} that we downloaded from the ENCODE database~\cite{sourceHMMdata}. This data set constitutes a list of start and stop positions associated with chromatin states called peaks. These peaks result from integrating several biological data sets, e.g.,  ChIP-seq and RNA-seq, with a multivariate hidden Markov model (HMM). The authors~\cite{hmm2011article} use $15$ ``HMM states'' (S1--S15): active promoter (S1), weak promoter (S2), inactive/poised promoter (S3), strong enhancer (S4 and S5), weak/poised enhancer (S6 and S7), insulator (S8), transcriptional transition (S9), transcriptional elongation (S10), weakly transcribed (S11), Polycomb-repressed (S12), heterochromatin (S13), and repetitive/copy number variation (S14 and S15).

The start and stop regions for these 15 HMM do not match perfectly with the Hi-C bins. To classify every Hi-C bin into one of these HMM states, we calculate the folds of enrichment (FE) relative to a chromosome-wide average according to the following steps.
\begin{enumerate}
    \item Count the number of peaks $k_X$ per bin, where $X=\rm{S1}, \dots, \rm{S15}$. Because some peaks span  multiple bins, we only count the peak starts.
    \item Calculate the peak frequency's expected value using the hypergeometric test (chromosome-wide sampling without replacement). The expected number of X peaks per bin is calculated as $\bar k'_X = K_X \times (n/N)$, where, $n$ is the number of peaks of any state in a bin, $N$ is the total number of peaks per chromosome, and $K_X$ is the total number of peaks for state $X$.
    \item Calculate the folds of enrichment $\mathrm{FE}_X$ for each HMM state $X$ per bin by dividing the observed by the expected peak number, $\mathrm{FE}_X = k_X/\bar k'_X$.
\end{enumerate}

 We note each Hi-C bin can be enriched in several chromatin states. Based on enrichment, we divide Hi-C bins into five groups (A--D) if $\mathrm{FE}_X > 1$:  
\begin{quote}
    \sffamily
    \begin{itemize}
    \item[(A)] Promoters: $X = $ S1 and S2.
    \item[(B)] Enhancers: $X = $ S4, S5, S6, and S7.
    \item[(C)] Transcribed regions: $X = $ S9, S10, and S11.
    \item[(D)] Heterochromatin and other repressive states: $X = $ S3, S13, S14, and S15.
    \item[(E)] Insulators: $X = $ S8
\end{itemize}

\end{quote}
Apart from these five groups, we assign bins that are not enriched in any state to the category "NA".

\section{Results}
\label{sec:results}

 \begin{figure}
     \includegraphics[width = \columnwidth]{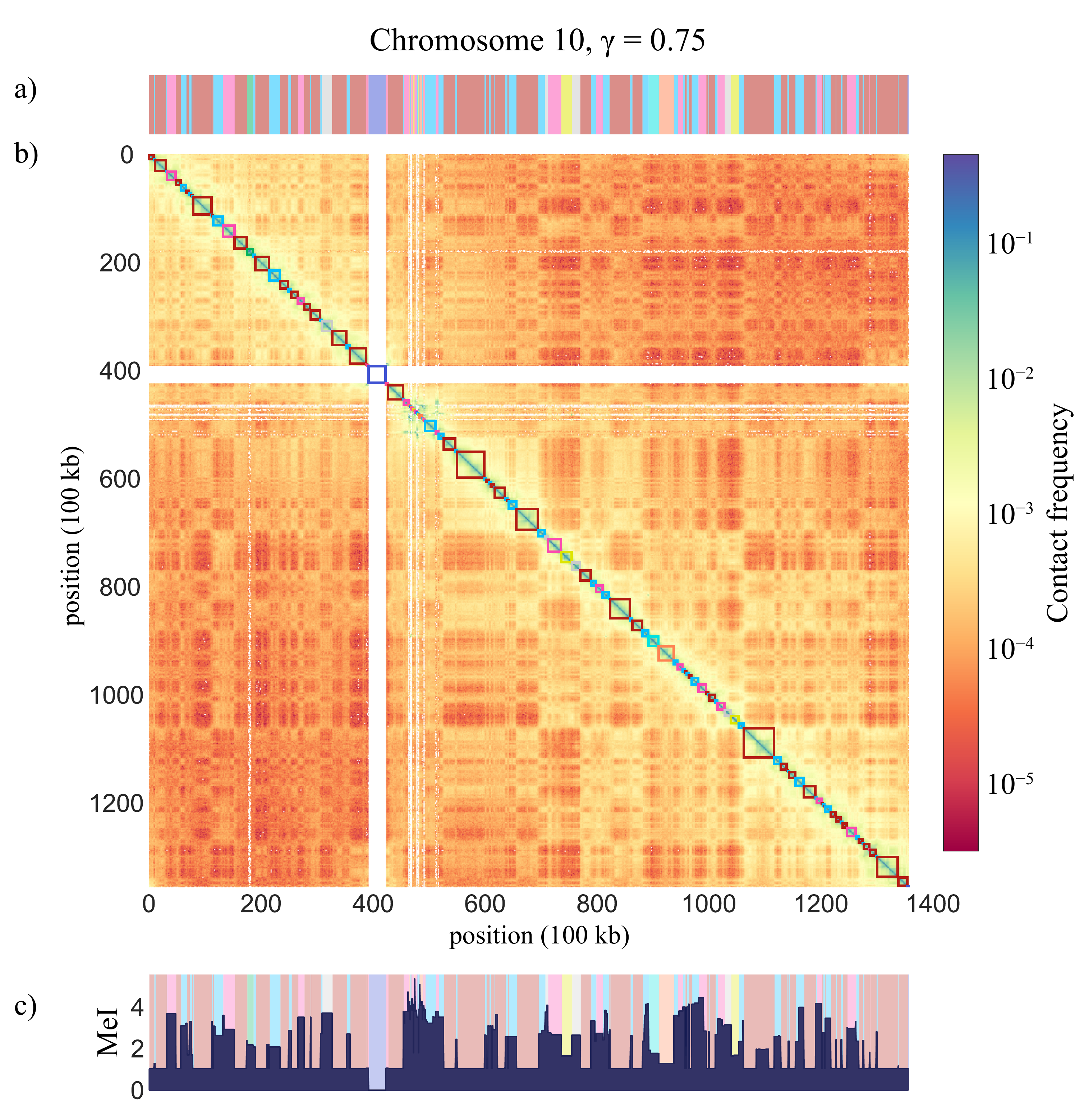}
     \caption{Membership inconsistency analysis for $\gamma=0.75$. The horizontal axis represents the genomic position along human chromosome 10.
     (a) Community membership of DNA segments shown as rectangles along the DNA sequence. Rectangles of the same color enclose DNA segments that are members of the same community (a single configuration).
     (b) Hi-C data is depicted as a heat map of contact frequencies between DNA-segment pairs. Rectangles along the main diagonal represent DNA segments' community membership from panel (a).
     (c) Nodes' membership inconsistency measured across 100 GenLouvain realizations. The dark-blue plot shows MeI scores of DNA segments in a chromosome (the network's nodes). The background of this panel shows the color-coded community membership from panel (a).}
     \label{fig:gL_on_HiC}
 \end{figure}

 \begin{figure*}
     \includegraphics[width = 18 cm]{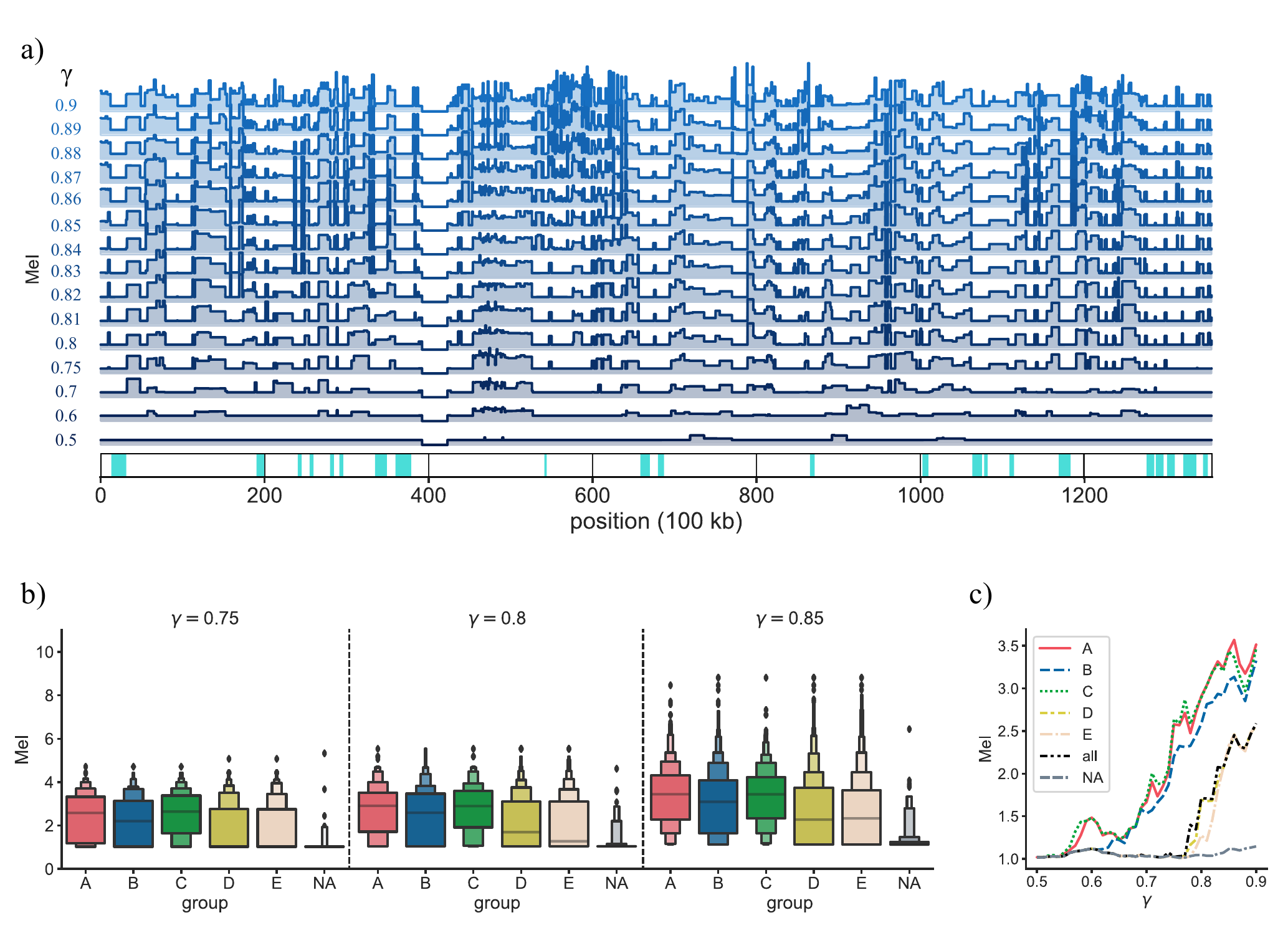}
     \caption{Membership inconsistency (MeI) across network scales ($0.5 \leq \gamma \leq 0.9$) on human chromosome 10. 
     (a) MeI profile for each resolution parameter $\gamma$. Below the waterfall plot, we draw blue squares that enclose DNA segments with low MeI scores across all scales.
     (b) Distribution of MeI scores shown for three network scales ($\gamma = 0.75, 0.80, 0.85$). We separated the distributions  into A,..., and E groups (and `NA') highlighting the nodes' chromatin state. Each distribution is shown as a boxen plot with a center line marking the median.
     (c) Median MeI across $\gamma$ and chromatin groups.
     }
     \label{fig:MeI_big_panel}
 \end{figure*}
 
\subsection{Local community inconsistency}
 
 We illustrated the local inconsistency associated with a single Hi-C map in Fig.~\ref{fig:gL_on_HiC}. This map depicts the number of contacts between all 100 kb DNA-segment pairs (KR normalized) in human chromosome 10. Along the diagonal, we highlight the GenLouvain-derived communities~\cite{GenLouvain}, at resolution parameter $\gamma = 0.75$. Squares sharing colors have the same community membership. These colors are better illustrated in the stripe above the map, showing how communities appear along the linear DNA sequence. We note that some scattered segments have the same color, which indicates that communities assemble DNA segments in 3D proximity, not only 1D adjacent neighbors. This contrasts the conventional notion of TADs, which comprise contiguous DNA stretches. To separate notations, we denote unbroken units of DNA stretches in a community as \emph{domains}.

It is essential to realize that the domains and communities in Figs.~\ref{fig:gL_on_HiC}a) and \ref{fig:gL_on_HiC}b) represent a single configuration, or partition, of Hi-C network communities at one specific resolution parameter value $\gamma$. Since GenLouvain uses a stochastic maximization algorithm, we expect to find other partitions if running it several times on the same data set, some of which may differ substantially. To quantify this variability, we generated $100$ independent network partitions and calculated the local inconsistency measure MeI~\cite{lee2021consistency} that quantifies how many different community configurations a single node effectively belongs to. We plot the MeI profile along chromosome 10 in Fig.~\ref{fig:gL_on_HiC}c). This profile shows that about half of the domains do not change community membership (the median value of MeI = 1.02), whereas the rest show significantly more variability (MeI~$\approx 4$). We also note that the MeI score is relatively uniform within each domain and that sharp MeI transitions occur near domain boundaries.

Based on previous work~\cite{lee2021consistency, CC1}, we anticipate that the MeI profile changes with the network scale. Therefore, we scanned through a wide range of community scales, extracted 3D communities, and calculated the MeI profile. We show the result from such a sweep in Fig.~\ref{fig:MeI_big_panel}a), where each MeI profile is associated with one $\gamma$ value. We note that some DNA regions have low MeI scores ($\gamma > 0.6$), which indicates that nodes in those regions mostly appear in the same communities for most $\gamma$ values. We indicate this as colored rectangles below the MeI profiles. But other DNA regions show the opposite behavior. These regions contain nodes that often do not appear in the same communities, which results in high and variable MeI values. Overall, the local node inconsistency grows as $\gamma$ becomes larger.

 \subsection{Local inconsistency and chromatin states}
 
To appreciate the MeI variations from a biological perspective,  we analyzed them relative to local chromatin states. As outlined in the Methods (Sec. \ref{sec:chromatin_states}), we use five states and calculate the folds of enrichment for each node. We denote the chromatin states as promoters (A), enhancers (B), transcribed regions (C), heterochromatin and other repressive states (D), and insulators (E).

Below the MeI profile in Fig.~\ref{fig:MeI_big_panel}, we show boxen plots for three $\gamma$ values. Each subplot illustrates the distribution of  MeI scores associated with each chromatin group (A--E); `NA' represents nodes that are not enriched in any chromatin type.  If following the MeI medians (horizontal lines), we note that groups A--C have consistently higher values than the rest. In panel (c), we explore this observation more thoroughly and plot the median MeI for several $\gamma$ values. The lines show that MeI grows with $\gamma$ and that groups A--C are more inconsistent than the chromosome-wide average (denoted `all'). This result suggests that nodes flagged as active chromatin have more variable node-community memberships.

\subsection{Cross-scale global inconsistency}
 
The previous subsection analyzed the cross-scale local inconsistency measure (MeI) for chromosome 10. Here, we extend the inconsistency analysis to all human chromosomes using the global inconsistency PaI instead of MeI (PaI yields one number per $\gamma$ value instead of a chromosome-wide profile). The PaI score measures the effective number of independent network partitions (see Sec.~\ref{sec:methods}). By the mathematical construction of PaI~\cite{lee2021consistency}, if there is no special scale of communities, the ``null-model'' behavior of PaI as a function of $\gamma$ would be as follows. As $\gamma$ gets larger from $\gamma = 0$, the PaI value start to increase as the average number of communities increases enough to form a certain level of inconsistency (for $\gamma = 0$, PaI trivially vanishes because there cannot be any inconsistency for the single community composed of all of the nodes). On the other extreme case of $\gamma \to \infty$, each individual node tends to form its own singleton community, so again there is no inconsistency, or PaI becomes zero. Therefore, if there is no particular characteristic scale of communities, the PaI curves against $\gamma$ would be single-peaked ones without any nontrivial behavior such as local minima. In reality, there are characteristic scales of communities, where PaI reach its local minima~\cite{lee2021consistency}, which indicate the most meaningful community scale. As our results show, the Hi-C communities also exhibit such characteristic scales. To better understand this metric, we revisit chromosome 10 before analyzing all human chromosomes.

\begin{figure}
     \includegraphics[width = \columnwidth]{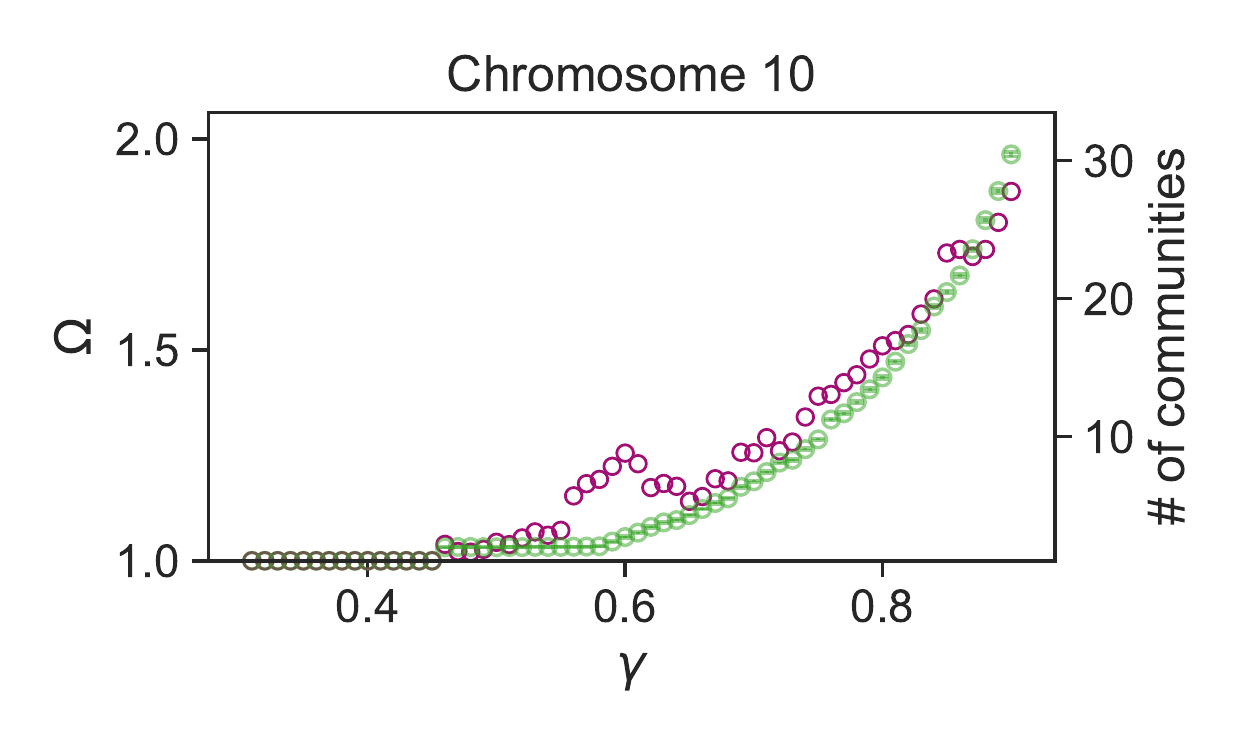}
     \caption{Global inconsistency measured by PaI (the violet circles) for human chromosome 10, along with the average number of communities (the green circles) across a range of $\gamma$ values with the small (smaller than the symbols themselves for all cases) error bars representing the standard deviation.}
     \label{fig:PaI_chr10}
 \end{figure}

We plot the PaI values for chromosome 10 in Fig.~\ref{fig:PaI_chr10} as violet circles.  When $\gamma$ is small ($<0.5$), we note that PaI has a plateau extending over several $\gamma$ values. Such a plateau is ideal for stable community partitions (see Fig.~\ref{fig:PaI}). However, this case is trivial because the community comprises the entire network (PaI = 1, thus one effective community). Next, if $\gamma$ increases above $0.5$, PaI starts to fluctuate, which indicates that partitions become more variable. Notably, the growing trend stops at $\gamma \approx 0.6$, and PaI decays to eventually reach a local minimum at $\gamma \approx 0.65$. This local minimum represents relatively stable communities, hinted by the small effective number of independent partitions at that scale. As we increase $\gamma$ above 0.65, the community structure becomes less and less stable, along with the rapidly growing number of communities (the green circles). However, asymptotically, the number of independent community ensembles grows slightly less than the number of communities per ensemble. For example, at $\gamma = 0.9$, there are $1.75$ independent ensembles (effective), each of which is composed of $25$ communities. As a final remark, it is essential to realize that the growing trend of PaI with $\gamma$ does not necessarily imply the lack of intrinsic organizational scales. Instead, it indicates fuzzy scale transitions where we observe a short range of stable communities at the local PaI minimum. 

 \begin{figure*}
     \includegraphics[width = \textwidth]{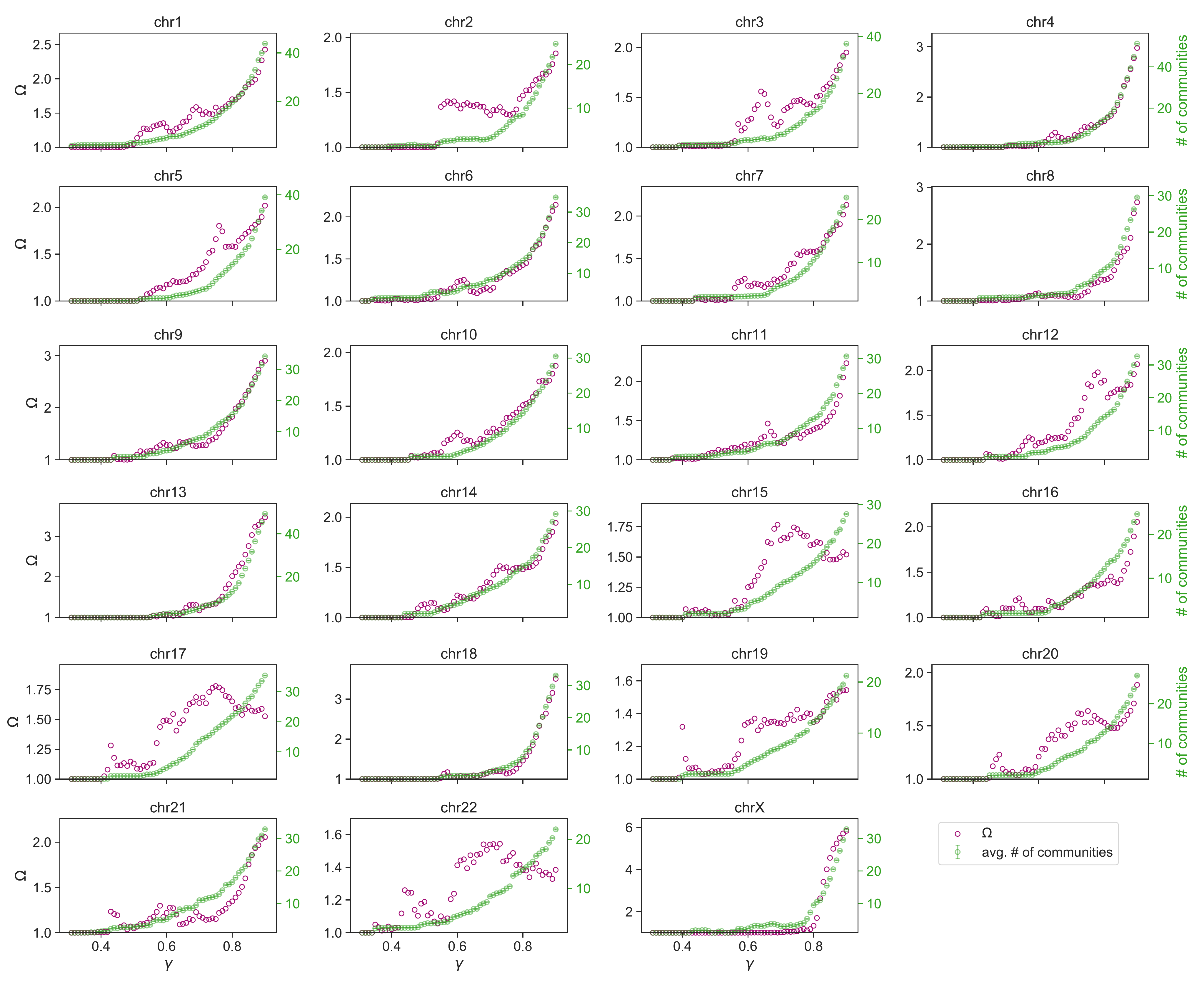}

     \caption{Chromosome-wide global community inconsistency measured by PaI. Each panel shows the values of the PaI metric (the violet circles) and the average number of communities (the green circles) across a range of $\gamma$ values with the small (smaller than the symbols themselves for all cases) error bars representing the standard deviation.}
     \label{fig:PaI_allchr}
\end{figure*}
 
When examining the PaI curve for chromosome 10, we noticed one significant local minimum with a relatively stable community partition. Next, we ask if similar inconsistency patterns appear across all human chromosomes. To this end, we plotted PaI against $\gamma$ for chromosomes 1--22 and X in Fig. \ref{fig:PaI_allchr}. 

We found several commonalities. First, similar to chromosome 10, most PaI curves have at least one local minimum and maximum. Some chromosomes even have two minima (e.g., chromosomes 1, 3, 9, 14, etc.), which indicate multiple stable scales of communities. Also, when $\gamma$ becomes large enough, the network enters the multi-community regime.
Second, as $\gamma$ grows, so does PaI. This growth indicates that community structures become increasingly inconsistent. Although it is natural to observe higher inconsistency for larger numbers of communities as there are more possible combinations. As future work, it would be informative to check its scaling behavior across different chromosomes and classify chromosomes based on the functional shapes of PaI and the number of communities.

\subsection{Node membership correlations}

 \begin{figure*}
     \includegraphics[width = \textwidth] {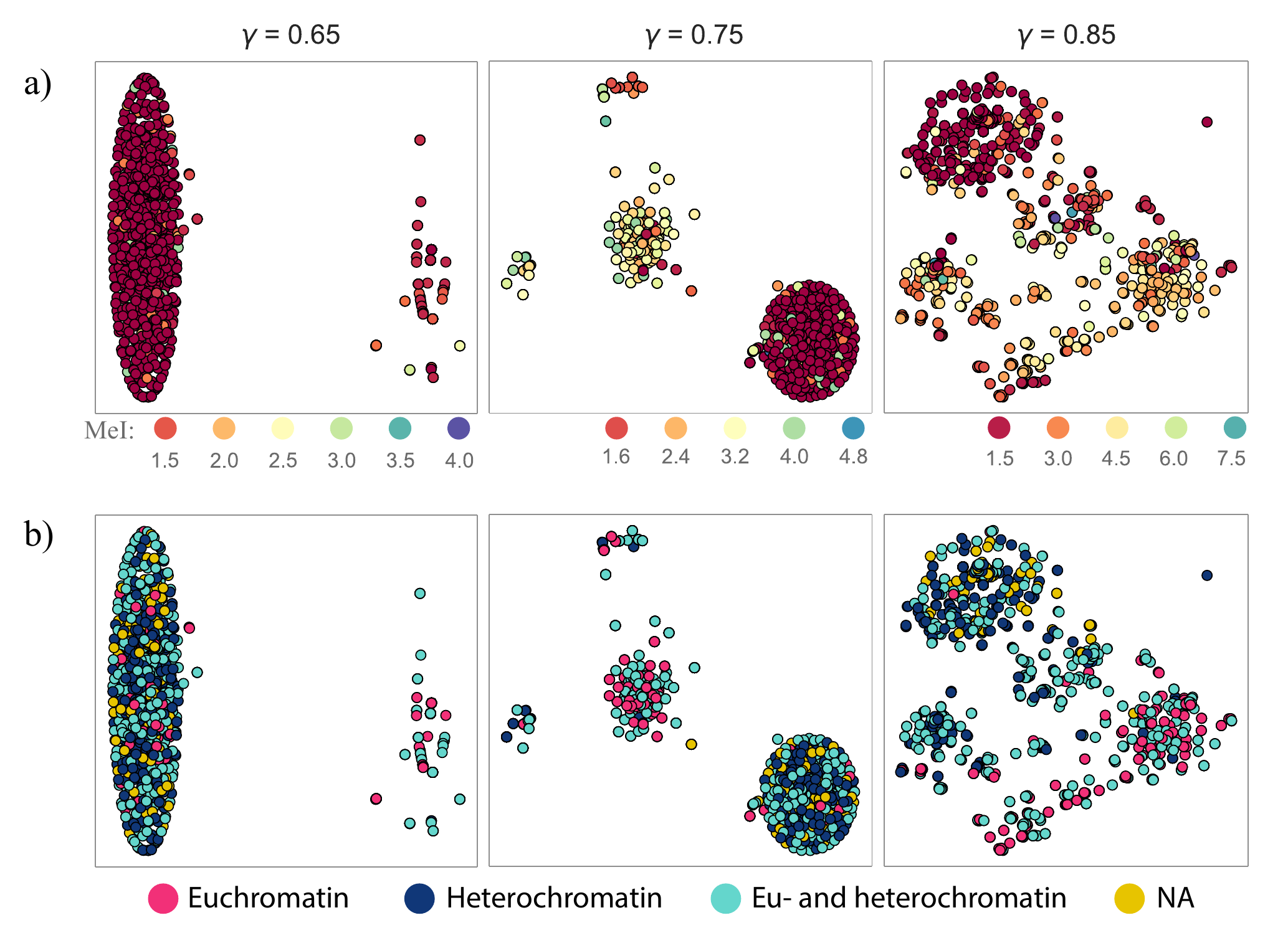}
     \caption{Node-community membership correlations across three resolution parameters ($\gamma = 0.65, 0.75, 0.85$) visualized with t-SNE dimension reduction. Nodes are represented as colored circles. Those having correlated community memberships over several GenLouvain realizations tend to cluster. In contrast, nodes that repel each other belong less likely to the same community.
     In (a), the colors indicate MeI scores (see legend under each plot). In (b) chromatin types (Eu- or Heterochromatin). 
     }
     \label{fig:tSNE}
 \end{figure*}

When analyzing the PaI and MeI scores across $\gamma$, we noted that the Hi-C networks exhibit relatively few independent partitions (e.g., PaI$_{\rm  chr10} < 3$), and that each node belongs to just a few communities (median MeI $ < 4$). This suggests that the community partitions are correlated. To better understand these correlations, we use a stochastic embedding technique called t-SNE that projects high-dimensional data clusters on a 2D plane (see Sec. \ref{sec:t_SNE}). In our case, the data set is the community membership per node over $100$ GenLouvain runs.

We show the t-SNE analysis in Fig.~\ref{fig:tSNE} for three $\gamma$ values, where each filled circle represents a network node (again, using chromosome 10). The closer two circles appear in the plot, the more correlated their node-community memberships are. As we increase $\gamma$, we note that node clusters split and that some circles become isolated. We interpret this as the ensemble of network partitions grows with $\gamma$ and becomes increasingly dissimilar.

While the clustering is identical in panels a) and b), we color-coded them differently to highlight specific features. In a), the colors represent the local inconsistency score (MeI). We note that nodes having high MeI tend to separate from nodes with low MeI. We also see that the low MeI nodes have relatively stronger correlations, thereby forming more distinct clusters. Panel a) also indicates that nodes with low MeI have similar node-community memberships.

In panel b), the color-coding illustrates the chromatin type. To simplify the analysis, we consider two large chromatin groups---Euchromatin and Heterochromatin---instead of the five we used before (Sec. \ref{sec:chromatin_states}. These two large groups reflect the traditional division into open and closed chromatin that is associated with active transcription and repression, respectively. In terms of our previous definitions, we form the following two groups: 
\begin{quote}
    \sffamily
    Euchromatin: A, B, and C, \\   
    Heterochromatin: D
\end{quote}
Note that we disregarded group E (``Insulators'') as it is associated with boundaries rather than long chromatin stretches, such as Eu- and Heterochromatin. Also, while most nodes belong to either Eu- or Heterochromatin, some are enriched in both types\footnote{Imagine a Venn diagram with two large circles portraying the chromatin enrichment for each node. While most nodes separate into either Eu- or Heterochromatin, the diagram shows a significant overlap where some nodes are enriched in both chromatin types. These nodes belong to the mixed group.}. 
We call this group ``mixed''. Finally, there is yet another node group that does not enrich any of the two chromatin types (``NA''). 

In panel b),  we observe that community membership correlations are associated with chromatin type. For example, when $\gamma=0.75$, Euchromatin and Mixed nodes separate from the large cluster and form new sub-groups. The nodes in these subgroups generally have high MeI scores indicating a larger variability in their community memberships.

Overall, panels (a) and (b) show a scale-dependent separation between communities associated with active and inactive DNA regions (red and blue nodes repel each other). This separation resembles the A/B compartmentalization but for small-scale 3D structures. Furthermore, the MeI score suggests that these structures have multiple independent ways to assemble if formed from Euchromatin nodes. This observation hints at higher structural variability of the accessible genome, which may reflect the dynamic nature of gene expression processes.

\section{Conclusions}
\label{sec:conclusions}

There is a growing awareness that Hi-C networks have a complex scale-dependent community structure. While some communities have a hierarchical or nested organization, others form a patchwork of partially overlapping communities. In this regard, Hi-C networks do not represent exceptions. Instead, they belong to the norm: most complex networks show convoluted multi-scale behaviors whenever competing organization principles shape the network structure. These principles force some nodes into ambiguous community memberships, making network partitioning challenging.

To better understand the scales where this might cause problems when clustering Hi-C data, we have analyzed the node-community variability over an ensemble of network partitions and estimated the ensemble's size. We have found that it typically grows as we zoom in to the network. However, this trend has significant breaks where the ensemble size drops at some specific network scales. This drop narrows the distribution of possible network partitions. We hypothesize that these minima represent the most common partitions of the average 3D  chromosome organization (over a cell population).

Moreover,  we have found nodes that belong to several communities when calculating the node-community membership variability. These ambiguous nodes act as bridges and are associated with specific chromatin types. For example, we have found the highest variability for nodes classified as enriched in active chromatin. This finding contrasts inactive (or repressed) chromatin nodes that typically exhibit a relatively more consistent community organization. One explanation is  Euchromatin's somewhat higher physical flexibility when exploring the nuclear 3D proximity in search of other DNA regions to form functional contacts (see a recent review \cite{misteli2020self-organizing-genome}). An alternative explanation is that fuzzy node-community memberships reflect significant cell-to-cell variations. While some physical interactions might be stable in one cell, they may be absent in another. Therefore, as Hi-C maps portray the average contact frequency over many cells, this variability may manifest in ambiguous nodes.

Finally, a recent study investigated the challenges in finding reliable communities in Hi-C data \cite{holmgren2022mapping}. This work aims to map out the landscape of feasible network partitions in Hi-C networks and found that the width of the landscape is scale-dependent. Our study takes a more node-centric view, where we calculated the local inconsistency of individual nodes and discovered that some nodes have fuzzy node-community memberships. Both studies highlight that finding reliable communities in Hi-C data is challenging, especially on some scales. One root cause is that Hi-C networks are almost entirely connected (with many weak links). Under these circumstances, we expect that Hi-C networks have several community divisions. Divisions that cannot be distinguished without additional data, such as gene expression or epigenetic profiles. This fundamental problem suggests that there is a significant likelihood of disagreement on the ideal network division between any community-finding or data-clustering methods. This challenge has likely contributed to debates on the actual differences between TADs and sub-TADs~\cite{dixon2016chromatin,eres2021tad}.

\acknowledgements
The authors thank Daekyung Lee for providing the Python codes to calculate the inconsistency measures~\cite{lee2021consistency}. S.H.L. was supported by the National Research Foundation (NRF) of Korea Grants Nos. NRF-2021R1C1C1004132 and NRF-2022R1A4A1030660. LL acknowledges financial support from the Swedish Research Council (Grant No. 2021-04080).

\bibliographystyle{unsrt}
\bibliography{refs}

\end{document}